# Theoretical study on the interfacial instability of a spherical droplet subject to vertical vibration

Yikai Li[1, *], Kun Wu[2, *], Dehua Liu[1] and Ru Xi[1]

1. School of Mechanical Engineering, Beijing Institute of Technology, Beijing 100081, China
2. State Key Laboratory of High Temperature Gas Dynamics, Institute of Mechanics, Chinese Academy of Sciences, Beijing 100190, China

**Abstract**

Interfacial instability would be aroused on a spherical liquid droplet when it is subject to external vertical vibration. In this paper, a linear analysis was conducted on this instability problem. The polar-angle dependent acceleration in the spherical coordinate is strongly coupled with the temporal and spatial component of the surface deformation displacement, which gives a recursion equation that implicitly expresses the dispersion relation between the growth rate and spherical mode numbers. The unstable regions (or unstable tongues) for the inviscid fluids considering latitudinal mode (longitudinal mode number $m = 0$) were derived and presented in the parameter plane. Compared with the solution of the spherical Faraday instability under radial vibration acceleration, the regions of harmonic unstable tongues for the mono-directional vibration case is much narrowed and the subharmonic unstable tongues almost become straight lines. The analysis shows that the latitudinal waves emerging on the spherical droplet surface ought to oscillate harmonically instead of subharmonically, which is opposite to the results for the case under radial vibration acceleration. A corresponding experiment of a liquid droplet lying on a vertically vibrating plate was conducted and the observations substantiate our theoretical predictions.

**Keywords**: Faraday instability, spherical droplet, vertical vibration, capillary waves

# 1. Introduction

Dynamics of a sessile liquid droplet lying on a vertically vibrating plate have been studied in many scientific and industrial fields, including spray coating **[1, 2]**, spray cooling **[3, 4]**, inkjet printing **[5, 6]**, crystallization **[7]** and humidification **[8]**. When the forcing vibration strength exceeds a certain threshold value, the liquid droplet becomes unstable to the external disturbance and the initial smooth surface wrinkles forming different patterns **[9-14]**. The amplitudes of the surface waves are enhanced with the further increase of forcing acceleration and breakup would occur at the tips of surface waves when the external acceleration increases over another threshold value **[15-22]**. These various phenomena have attracted so many academic attentions because they involve enormous fundamental interesting physics, such as the interface dynamics, pattern selection and liquid atomization.

One focus in the previous studies is the deformation and macroscopic movement of the droplet under different vibrating conditions and different surface conditions of the supporting solid plate. Noblin et al. **[23]** experimentally studied the effects of vertical vibration on the non-wetting sessile liquid droplet. Droplet deformations with tuning the frequency and displacement amplitude of the vibration were recorded. Two types of oscillations were categorized according to the different behaviors of contact line movements under different vibration amplitudes. Mettu and Chaudhury **[24]** studied the resonance frequencies of small sessile liquid droplets on both hydrophilic and superhydrophobic plates subjected to a white noise vibration. Different resonant modes were identified in experiments as functions of the droplet mass, contact angle, and liquid properties, which can be utilized as a novel way to measure the surface tension coefficient and viscosity for very small droplets. To overcome capillary forces and contact line retention forces in the microfluidic droplet application field, surface acoustic waves (SAW) are devised to manipulate the droplets. Miyamoto et al. **[25]** experimentally determined two nonlinear phenomena for droplet vibration under SAW excitation, *i.e.*, the resonance frequency decreases with the increase of SAW input and multi-frequencies vibration of waves on the droplet surface, besides the fundamental one, can be observed. Baudoin et al. **[26]** studied the low power sessile droplets actuation by modulated SAW and showed that the acoustic power required to move or deform droplets is greatly reduced through modulating the acoustic signal around Rayleigh-Lamb inertio-capillary frequencies. Electronic power has also been employed as a source to vibrate the liquid droplet lying on a plate. Yamada et al. **[27]** experimentally investigated the vibration of a water droplet located on the surface of a hydrophobic plate under an alternative current (AC) electric field of direction parallel to the plate surface. They have confirmed the effect of the surface property of hydrophobic materials on the resonance phenomena of the water droplet.

The surfaces of liquid bulks, including liquid layer and droplet, exhibit various standing wave patterns when they are subjected to vertical vibrations, which is another focus of the interfacial dynamics induced by external vibration. The oscillation frequency of surface wave patterns under external vibration was first theoretically analyzed by Benjamin and Ursell **[28]**, who explained why the observed frequency of surface waves can be half and synchronized with the external vibration in experiments. They derived the standard Mathieu equation for the inviscid fluids to govern the linear growth of surface deformation amplitude. The damping effects due to viscosity was then incorporated by Kumar **[29]** in their theoretical works, which provided the threshold acceleration to trigger the instability for viscous liquids. To predict different pattern selections of surface waves, nonlinear effects must be considered **[10]**. By means of quasi-potential approximation and multiscale asymptotic expansion, Zhang and Viñals **[12]** presented a theoretical study on the nonlinear pattern formation mechanism for low-viscosity fluids. They claimed that three-wave resonant interactions are important for the pattern selection. Chen and Viñals **[9]** derived a standing wave amplitude equation from Navier-Stokes equations without the assumption of low viscosity. Predictions of different selected patterns as a function of fluid viscosity and forcing frequency were presented. Pattern forming dynamics on the liquid surface with external vibration of multiple frequencies were also treated theoretically **[13, 30]** to explain the pattern selection in relevant experiments. Linear analysis on the instability of a levitated spherical droplet with radial vibration **[31, 32]** and numerical simulation of the wave patterns on the droplet surface **[14]** were not conducted until recent time.

The vertical vibration of a plate can also be employed to realize the breakup and atomization of a bulk of liquid lying on it if the vibration amplitude is sufficiently large, whereby many atomization techniques have been proposed. Qi et al. **[33]** exposed a millimeter-size liquid droplet lying on a single-crystal lithium niobate piezoelectric substrate to the SAW to realize atomization of the parent droplet. The fountain of aerosol droplets is not perpendicular to the substrate due to leakage of acoustic radiation at the Rayleigh angle. Through scaling theory and simple numerical modeling, they elucidated the mechanism of aerosol droplet formation as the interaction among viscous, capillary and inertial forces. A liquid droplet placed on a vibrating diaphragm can burst into a fine spray of smaller droplets, which is called vibration induced droplet atomization (VIDA) by James et al. **[17, 18, 20]**, who have systematically studied the basic physics of bursting dynamics for this variable mass system. Through a numerical simulation **[18]**, they obtained an acceleration threshold of vibration for low-frequency single-droplet ejection, which coincides with the one for the rare droplet ejection from a liquid layer subject to vertical vibrations **[15, 16]**. The threshold condition for the dense droplet ejections to form a spray was proposed by Li and Umemura **[21]** after numerically studied the detailed microscopic mechanism of the ligament formation from the liquid surface under vertical vibration **[19]**. When the frequency of forcing vibration increases to ultrasonic range (>20 kHz), so-called ultrasonic atomization will be realized. Lang **[34]** experimentally measured the mean size of droplets for a wide range of ultrasonic frequencies and proposed an equation correlated the mean size of atomized droplets, frequency and liquid properties. Avvaru et al. **[35]** then experimentally studied the role of liquid viscosity and non-Newtonian effect played in the droplet size distribution. Gaete-Garretón et al. **[36]** determined the displacement amplitude threshold for atomization at different frequencies by fitting the experimental data from 5 kHz to 50 kHz.

In the present study, the problem of VIDA is revisited, with the focus on the theoretical analysis on the interfacial instability of a spherical droplet under mono-directional vibration, which, to the best of our knowledge, has not yet been studied in the previous literatures. The most-related research is the linear analysis on the spherical Faraday instability with radial parametric acceleration conducted by Adou and Tuckerman **[31]** recently. Different from their study, the direction of acceleration acting on the droplet in the present consideration is vertical rather than radial, which makes the problem much more complicated because the radial component of acceleration varies at different polar angles. This polar-angle dependence of radial acceleration will cause more mathematical complications in the spherical coordinate and the solution will show more physics associated with the vibration induced interfacial instability. As will be shown below, the acceleration is strongly coupled with the temporal and spatial component of the surface deformation displacement, which would cause the resultant response of unstable waves on the droplet surface to show different dynamics from that under radial vibrating accelerations.

The remaining parts of this paper are organized as follows. In §2, the mathematical formulation for the problem considered in the present study is described, with the general solutions being derived in §3. In §4, the instability diagrams for the inviscid fluids with different mode numbers are compared with those of the radial acceleration conditions. Then, the response of the excited unstable surface waves on a liquid droplet lying on a vertically vibrating plate are experimentally recorded and discussed in §5 to verify the theoretical predictions. The whole contents are finally summarized in §6.

## 2. Mathematical derivation

The physical model we considered in the present study is shown as Fig. 1. A perfect hemispherical liquid droplet with the radius of $R_0$ is initially placed on a solid plate which oscillates vertically with the angular frequency $\Omega$ ($\Omega = 2\pi f$, where $f$ is the oscillation frequency) and displacement amplitude $\Delta_0$. The effects of the gravity and contact angle are neglected since they play secondary roles in the unstable capillary waves caused by the external vibration with relatively high frequency, which is the major focus of the present study. The inviscid gas surrounding the liquid droplet is quiescent initially and both the liquid and gas phases are assumed to be incompressible. The densities of the liquid and gas phases are denoted as $\rho_1$ and $\rho_2$, respectively. For typical cases of the unstable waves excited by vertical vibration on the droplet surface, the density of surrounding gas is negligibly small compared with that of liquid droplet ($\rho_1 \gg \rho_2$). The coefficient of surface tension σ acting on the liquid-gas interface is assumed to be constant. Furthermore, the wetting effect of the bottom solid plate is neglected in the present study.

In the reference frame attached to the vibrating plate, the vertical vibration can be considered as a body force acting on the liquid droplet. In the present study, we focus on the interfacial dynamics of the droplet subject to an infinitesimal initial disturbance in the linear regime, which indicates the governing equations for the motion of both phases are

$$\nabla \cdot \boldsymbol{u}_i = 0 \tag{1}$$

$$\frac{\partial \boldsymbol{u}_i}{\partial t} = -\frac{1}{\rho_i}\nabla p_i + \nu_i \nabla^2 \boldsymbol{u}_i + \boldsymbol{A} \tag{2}$$

where $\boldsymbol{A} = A_0 sin(\Omega t)cos\theta \boldsymbol{e}_r - A_0 sin(\Omega t)sin\theta \boldsymbol{e}_\theta = \nabla[A_0 sin(\Omega t)rcos\theta]$ is the acceleration vector acting on the liquid droplet and evolving sinusoidally over time $t$, with the acceleration amplitude $A_0 = \Delta_0 \Omega^2$, $\boldsymbol{e}_r$ and $\boldsymbol{e}_\theta$ being the unit vector along the radial and polar direction in the spherical coordinate system $(r, \theta, \varphi)$ as shown in Fig. 1; the subscripts $i = 1, 2$ represent the physical quantities for the liquid inside and gas outside the droplet, respectively; $\boldsymbol{u}_i$ is the perturbed velocity vector, $p_i$ the pressure, $\nu_i = \mu_i/\rho_i$ the kinematic viscosity ($\nu_2 = 0$ since the gas viscosity plays relatively subordinate role in this problem).

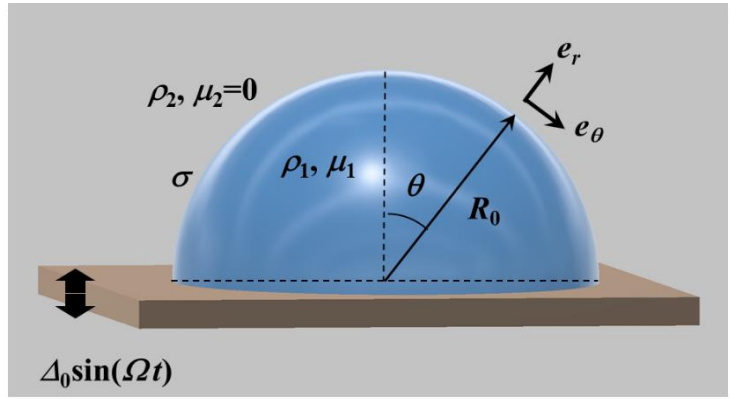

Fig. 1. Schematic of the physical model for the interfacial instability of a semi-spherical droplet subject to vertical vibration.

The linearized kinematic condition on the droplet surface is given by considering the continuous nature of the normal velocity across the interface, which can be expressed as

$$\frac{\partial \eta}{\partial t} = u_{1r}|_{r=R_0+\eta} = u_{2r}|_{r=R_0+\eta} \qquad (3)$$

where $\eta = \eta(\theta,\varphi,t)$ is the displacement of deformation on the droplet surface and $u_{ir}$ $(i=1,2)$ is the radial component of the velocity vector $\boldsymbol{u}_i$.

The stresses acting normal to the interface from liquid and gas sides should be balanced by satisfying [37]

$$p_1|_{r=R_0+\eta} - p_2|_{r=R_0+\eta} = \frac{2\sigma}{R_0} - \frac{\sigma}{R_0^2}(2\eta + \nabla_H^2 \eta) + 2\mu_1 \frac{\partial u_{1r}}{\partial r}\bigg|_{r=R_0+\eta} \qquad (4)$$

where

$$\nabla_H^2 = \frac{1}{\sin\theta}\frac{\partial}{\partial\theta}\left(\sin\theta\frac{\partial}{\partial\theta}\right) + \frac{1}{\sin^2\theta}\frac{\partial^2}{\partial\varphi^2} \qquad (5)$$

is the horizontal spherical Laplacian operator. The stresses acting tangential to the interface from opposite sides should satisfy

$$\varepsilon_{r\theta} = \mu_1\left(\frac{1}{r}\frac{\partial u_{1r}}{\partial \theta} - \frac{u_{1\theta}}{r} + \frac{\partial u_{1\theta}}{\partial r}\right) = 0 \qquad (6)$$

and

$$\varepsilon_{r\varphi} = \mu_1\left(\frac{1}{r\sin\theta}\frac{\partial u_{1r}}{\partial \varphi} - \frac{u_{1\varphi}}{r} + \frac{\partial u_{1\varphi}}{\partial r}\right) = 0 \qquad (7)$$

where $\varepsilon_{r\theta}$ and $\varepsilon_{r\varphi}$ are the tangential stresses along polar and azimuthal directions, respectively.

To solve the linearized VIDA problem in the spherical coordinate system as described above, it is more convenient to expand the displacement disturbance on the droplet surface $\eta = \eta(\theta,\varphi,t)$ in series of spherical harmonics $Y_l^m(\theta,\varphi) = P_l^m(\cos\theta)e^{im\varphi}$ as

$$\eta(\theta,\varphi,t) = \sum_{l=1}^{+\infty}\sum_{m=-l}^{l} \eta_l^m(t) Y_l^m(\theta,\varphi) \qquad (8)$$

Considering that the source term of the momentum equation (2) evolves over time, we shall write the time-dependent coefficient in Eq. (8) in the Floquet form [29, 38]:

$$\eta_l^m(t) = e^{(\beta+i\gamma)t} \sum_n \eta_n(l,m) e^{in\Omega t} = \sum_n \eta_n(l,m) e^{\zeta_n t} \tag{9}$$

where $\beta + i\gamma$ is the Floquet exponent, $\eta_n$ the coefficient of Fourier mode $n$, and $\beta$, the real part of $\zeta_n = \beta + i(\gamma + n\Omega)$, can be regarded as the growth rate for a spherical mode composite $(l,m)$. In the following, we shall derive the solution to Eqs. (1)-(9) to find the condition for the diverging/decaying growth of the deformation on the droplet surface.

## 3. General Solutions

The basic strategy to solve the eigenvalue problem governed by Eqs. (1)-(9) as described above is similar to our previous one to study the spherical Faraday instability subjected to radial acceleration [32]. Since the direction of acceleration acting on the droplet is vertical rather than radial in the VIDA problem, the mathematical derivation is much more complicated, which will be demonstrated as follows.

The solution for the liquid phase inside the droplet is first to be derived. Using the equation $\nabla \times \nabla \times \boldsymbol{u} = \nabla(\nabla \cdot \boldsymbol{u}) - \nabla^2 \boldsymbol{u}$ along with the continuity equation (1), we can easily reduce the momentum equation (2) into

$$\frac{\partial \boldsymbol{u}_1}{\partial t} = -\frac{1}{\rho_1}\nabla p_1 - \nu_1 \nabla^2 \times \boldsymbol{u}_1 + \boldsymbol{A} \tag{10}$$

The velocity vector $\boldsymbol{u}_1$ can be split in the irrotational component $\nabla\phi_1$ and the rotational component $\boldsymbol{\psi}_1$, i.e., $\boldsymbol{u}_1 = \nabla\phi_1 + \boldsymbol{\psi}_1$, which follow

$$\nabla^2 \phi_1 = 0 \tag{11}$$

$$p_1 = -\rho_1 \frac{\partial \phi_1}{\partial t} + \rho_1 A_0 \cos(\Omega t) r\cos\theta + C_1(\theta, t) \tag{12}$$

$$\nabla \cdot \boldsymbol{\psi}_1 = 0 \tag{13}$$

$$\frac{\partial \boldsymbol{\psi}_1}{\partial t} + \nu_1 \nabla^2 \times \boldsymbol{\psi}_1 = 0 \tag{14}$$

according to Eq. (10), where $C_1(\theta, t)$ is an integral constant as a function of $\theta$ and $t$.

Considering that the velocity potential $\phi_1$ should be a finite value at $r = 0$, the general solution to Eq. (11) in the spherical coordinate should be

$$\phi_1(r,\theta,\varphi,t) = \sum_n B_n e^{\zeta_n t} r^l Y_l^m(\theta,\varphi) \qquad (15)$$

for each spherical harmonic $Y_l^m(\theta,\varphi)$, where $B_n$ is the coefficient to be determined through the boundary conditions. It should be noticed that the indices $(l,m)$ and corresponding sums for $\phi_1$ and other similar quantities are omitted for mathematical conciseness in Eq. (15) and in the following derivations.

As for the rotational component $\boldsymbol{\psi}_1$ governed by Eqs. (13) and (14), we followed the method proposed by Chandrasekhar [39], which gives the three components of $\boldsymbol{\psi}_1$ as

$$\begin{cases} \psi_{1r}(r,\theta,\varphi,t) = \sum_n \dfrac{l(l+1)}{r^2} \Psi_1(r) Y_l^m(\theta,\varphi) e^{\zeta_n t} \\ \psi_{1\theta}(r,\theta,\varphi,t) = \sum_n \dfrac{1}{r} \dfrac{d\Psi_1(r)}{dr} \dfrac{\partial Y_l^m(\theta,\varphi)}{\partial \theta} e^{\zeta_n t} \\ \psi_{1\varphi}(r,\theta,\varphi,t) = \sum_n \dfrac{1}{r\sin\theta} \dfrac{d\Psi_1(r)}{dr} \dfrac{\partial Y_l^m(\theta,\varphi)}{\partial \varphi} e^{\zeta_n t} \end{cases} \qquad (16)$$

with $\Psi_1(r) = D_n r^{1/2} J_{l+1/2}(is_n r)$, where $D_n$ is another coefficient to be determined, $J_{l+1/2}$ is the spherical Bessel function of order $l+1/2$ and $s_n = \sqrt{\zeta_n/\nu_1}$.

Substituting Eqs. (15) and (16) into the boundary conditions (3), (6) and (7), we can obtain the expressions for the coefficients $B_n$ and $D_n$ as follows:

$$B_n = \frac{\eta_n \zeta_n}{l r_0^{l-1}} \left[ 1 + \frac{2(l^2-1)}{2x Q_{l+1/2}(x) - x^2} \right]$$

$$D_n = -\frac{2(l-1)\eta_n \zeta_n r_0^{3/2}}{l \left[ 2x J_{l+3/2}(x) - x^2 J_{l+1/2}(x) \right]} \qquad (17)$$

where $Q_{l+1/2}(x) = J_{l+3/2}(x)/J_{l+1/2}(x)$ and $x = is_n r_0$.

Besides the presence of the time-dependent term $\cos(\Omega t)$ in Eq. (12) which makes the equation inhomogeneous in time, the existence of spatial dependent term $\cos\theta$ additionally makes the equation inhomogeneous in space. This would result in complicated coupling between different Fourier modes $n$ and different spherical modes $l$ considering the Floquet form of the surface disturbance $\eta$ as Eqs. (8) and (9), which gives

$$\begin{aligned}
\cos(\Omega t)\cos\theta \cdot \eta &= \cos(\Omega t)\cos\theta \cdot \sum_{l=1}^{+\infty}\sum_{m=-l}^{l}\eta_l^m(t)Y_l^m(\theta,\varphi) \\
&= \frac{e^{i\Omega t}+e^{-i\Omega t}}{2}\left\{\sum_{l=1}^{+\infty}\sum_{m=-l}^{l}\frac{l-m}{2l-1}Y_l^m(\theta,\varphi)\eta_{l-1}^m(t) + \sum_{l=1}^{+\infty}\sum_{m=-l}^{l}\frac{l+m+1}{2l+3}Y_l^m(\theta,\varphi)\eta_{l-1}^m(t)\right\} \\
&= \frac{e^{i\Omega t}+e^{-i\Omega t}}{2}\left\{\sum_{l=1}^{+\infty}\sum_{m=-l}^{l}\frac{l-m}{2l-1}Y_l^m(\theta,\varphi)\left[e^{(\beta+i\gamma)t}\sum_n \eta_n(l-1,m)e^{in\Omega t}\right]\right. \\
&\quad\left. + \sum_{l=1}^{+\infty}\sum_{m=-l}^{l}\frac{l+m+1}{2l+3}Y_l^m(\theta,\varphi)\left[e^{(\beta+i\gamma)t}\sum_n \eta_n(l+1,m)e^{in\Omega t}\right]\right\} \\
&= \frac{1}{2}\left\{\sum_{l=1}^{+\infty}\sum_{m=-l}^{l}\frac{l-m}{2l-1}Y_l^m(\theta,\varphi)\sum_n[\eta_{n-1}(l-1,m)+\eta_{n+1}(l-1,m)]e^{\zeta_n t}\right. \\
&\quad + \sum_{l=1}^{+\infty}\sum_{m=-l}^{l}\frac{l+m+1}{2l+3}Y_l^m(\theta,\varphi)\sum_n[\eta_{n-1}(l+1,m) \\
&\quad\left. + \eta_{n+1}(l+1,m)]e^{\zeta_n t}\right\}
\end{aligned} \quad (18)$$

using the relation

$$\cos\theta P_l^m(\cos\theta) = \frac{l-m+1}{2l+1}P_{l+1}^m(\cos\theta) + \frac{l-m}{2l+1}P_{l-1}^m(\cos\theta) \quad (19)$$

With the solution of the velocity potential from Eqs. (15) and (17), and the recurrence relation (12), the pressure on the inner side of the droplet surface can be obtained:

$$\begin{aligned}
p_1|_{r=R_0+\eta} &= -\sum_{l=0}^{+\infty}\sum_{m=0}^{l}\sum_{n=0}^{+\infty}\frac{\rho_1\eta_n(l,m)\zeta_n^2 r_0}{l}\left[1+\frac{2(l^2-1)}{2xQ_{l+1/2}(x)-x^2}\right]e^{\zeta_n t}Y_l^m \\
&\quad + \rho_1 A_0\cos(\Omega t)\cos\theta r_0 + C_1 \\
&\quad + \frac{1}{2}\rho_1 A_0\sum_{l=0}^{+\infty}\sum_{m=0}^{l}\sum_{n=0}^{+\infty}\left\{\frac{l-m}{2l-1}[\eta_{n-1}(l-1,m)+\eta_{n+1}(l-1,m)]\right. \\
&\quad\left. + \frac{l+m+1}{2l+3}[\eta_{n-1}(l+1,m)+\eta_{n+1}(l+1,m)]\right\}e^{\zeta_n t}Y_l^m
\end{aligned} \quad (20)$$

For the surrounding inviscid gas phase which is initially stagnant and irrotational, a velocity potential $\phi_2$ satisfying $\boldsymbol{u}_2 = \nabla\phi_2$ is introduced and the continuity equation (1) reduces to

$$\nabla^2\phi_2 = 0 \tag{21}$$

Integrating the momentum equation for the gas phase gives

$$p_2 = -\rho_2\frac{\partial\phi_2}{\partial t} + \rho_2 A_0 \cos(\Omega t) r\cos\theta + C_2(\theta, t) \tag{22}$$

where $C_2(\theta, t)$ is another integral constant as a function of $\theta$ and $t$.

Eq. (21) can be solved by considering the linearized kinematic boundary condition (3) and the fact that the velocity at $r \to \infty$ cannot be diverge, which gives

$$\phi_2 = -\sum_n \eta_n \frac{\zeta_n r_0}{l+1}\left(\frac{r}{r_0}\right)^{-(l+1)} e^{\zeta_n t} Y_l^m \tag{23}$$

for each spherical harmonic $Y_l^m(l, m)$. With Eq. (23) and the recurrence relation (18), the pressure on the outer side of the droplet surface can be obtained:

$$\begin{aligned}
p_2|_{r=R_0+\eta} = &-\sum_{l=0}^{+\infty}\sum_{m=0}^{l}\sum_{n=0}^{+\infty}\frac{\rho_2\eta_n(l,m)\zeta_n^2 r_0}{l} e^{\zeta_n t} Y_l^m + \rho_2 A_0\cos(\Omega t)\cos\theta r_0 + C_2 \\
&+ \frac{1}{2}\rho_2 A_0 \sum_{l=0}^{+\infty}\sum_{m=0}^{l}\sum_{n=0}^{+\infty}\Big\{\frac{l-m}{2l-1}[\eta_{n-1}(l-1,m) + \eta_{n+1}(l-1,m)] \\
&\quad + \frac{l+m+1}{2l+3}[\eta_{n-1}(l+1,m) + \eta_{n+1}(l+1,m)]\Big\} e^{\zeta_n t} Y_l^m
\end{aligned} \tag{24}$$

Substituting solutions (20) and (24) into the pressure balance condition (4) and with the identity

$$\Delta_{\theta\varphi} Y_l^m = \nabla_H^2 Y_l^m = -l(l+1)Y_l^m \tag{25}$$

we have

$$\sum_{l=0}^{+\infty}\sum_{m=0}^{l}\sum_{n=0}^{+\infty} e^{\zeta_n t} Y_l^m \left\{ \left[ -\left(\frac{\rho_1}{l} + \frac{\rho_2}{l+1}\right) r_0 \zeta_n^2 \right.\right.$$
$$+ 2\mu_1 \zeta_n \frac{l-1}{lr_0} \frac{(2l+1)x - 2l(l+2)Q_{l+1/2}(x)}{2xQ_{l+1/2}(x) - x^2}$$
$$\left.- \frac{\alpha(l-1)(l+2)}{r_0^2} \right] \eta_n(l,m) \quad (26)$$
$$+ \frac{1}{2}(\rho_1 - \rho_2) A_0 \left[ \frac{l-m}{2l-1} (\eta_{n-1}(l-1,m) + \eta_{n+1}(l-1,m)) \right.$$
$$\left.\left.+ \frac{l+m+1}{2l+3} (\eta_{n-1}(l+1,m) + \eta_{n+1}(l+1,m)) \right] \right\}$$
$$+ (\rho_1 - \rho_2) A_0 \cos(\Omega t) r_0 \cos\theta + C_1(\theta, t) - C_2(\theta, t) - \frac{2\alpha}{r_0} = 0$$

This equation requires two conditions to be valid. The first one is

$$(\rho_1 - \rho_2) A_0 \cos(\Omega t) r_0 \cos\theta + C_1(\theta, t) - C_2(\theta, t) - \frac{2\alpha}{r_0} = 0 \quad (27)$$

which can used to determine the integral constants $C_1$ and $C_2$ with an additional boundary condition of $p_1$ at $r = 0$. The second condition is that the coefficients of the Fourier term in Eq. (26) should satisfy

$$\frac{1}{2}(\rho_1 - \rho_2) A_0 \left[ \frac{l-m}{2l-1} (\eta_{n-1}(l-1,m) + \eta_{n+1}(l-1,m)) \right.$$
$$\left.+ \frac{l+m+1}{2l+3} (\eta_{n-1}(l+1,m) + \eta_{n+1}(l+1,m)) \right]$$
$$= \left[ \left(\frac{\rho_1}{l} + \frac{\rho_2}{l+1}\right) r_0 \zeta_n^2 \right. \quad (28)$$
$$- 2\mu_1 \zeta_n \frac{l-1}{lr_0} \frac{(2l+1)x - 2l(l+2)Q_{l+1/2}(x)}{2xQ_{l+1/2}(x) - x^2}$$
$$\left.+ \frac{\alpha(l-1)(l+2)}{r_0^2} \right] \eta_n(l,m)$$

because different Fourier modes are orthogonal. Eq. (28) is a recursion equation about the Fourier components, which implicitly expresses the dispersion relation between the growth rate and spherical mode numbers. Note that different from the previous theoretical results for the spherical Faraday instability under radial acceleration **[31, 32]**, the Fourier modes from different spherical harmonics are closely coupled in Eq. (28), which makes the problem more complicated. Theoretical analyses derived from Eq. (28) and their inspiration for the liquid droplet instability will be detailed in the following sections.

# 4. Instability diagrams for the spherical droplet subject to vertical vibration

For typical conditions of the unstable waves excited on the liquid (such as water) droplet, the acceleration of external vibration is sufficiently large to overcome the stabilizing effects played by capillarity **[17]**, and the effects of liquid viscosity are overwhelmed by inertial force. It is reasonable to first neglect the viscous effect to emphasize the major dynamics associated with the interfacial instability caused by vertical vibration in the present study. Furthermore, the liquid density is far larger than the gas density in most application fields of atomization, so the gas density can also be neglected.

For the inviscid liquid droplet immersed in a gaseous atmosphere, the recursion equation (28) can be reduced to

$$\frac{1}{2}\rho_1 A_0 \left[ \frac{l-m}{2l-1}\left(\eta_{n-1}(l-1,m) + \eta_{n+1}(l-1,m)\right) \right.$$
$$\left. + \frac{l+m+1}{2l+3}\left(\eta_{n-1}(l+1,m) + \eta_{n+1}(l+1,m)\right) \right] \quad (29)$$
$$= \left[ \frac{\rho_1}{l}r_0\zeta_n^2 + \frac{\alpha(l-1)(l+2)}{r_0^2} \right]\eta_n(l,m)$$

For an instability problem, the first concern is the boundary or critical values between the stable and unstable behaviors. In the previous studies on the spherical Faraday instability with the radial acceleration, this boundary can be obtained by solving the eigenvalues of a coefficient matrix with infinite dimensions **[31, 32]**. The similar philosophy is employed in the present study. To obtain the instability boundary, we shall set the growth rate of each mode $\beta$ to be zero. As a result, the recursion equation (29) is transformed to

$$\frac{l}{2[(l-1)(l+2)l - (\Omega/\Omega_c)^2(\hat{\gamma}+n)^2]}\left[ \frac{l-m}{2l-1}\left(\eta_{n-1}(l-1,m)\right.\right.$$
$$\left. + \eta_{n+1}(l-1,m)\right)$$
$$\left. + \frac{l+m+1}{2l+3}\left(\eta_{n-1}(l+1,m) + \eta_{n+1}(l+1,m)\right) \right] \quad (30)$$
$$= \frac{1}{A_0/A_c}\eta_n(l,m)$$

with the definitions of

$$\Omega_c^2 = \frac{\alpha}{\rho_1 r_0^3} \quad and \quad A_c = r_0 \Omega_c^2 = \frac{\alpha}{\rho_1 r_0^2} \tag{31}$$

which are two characteristic parameters determined by the fluid property and the size of droplet. $(\Omega/\Omega_c)$ and $A_0/A_c$ are two dimensionless parameters to characterize a prescribed experimental realization. It is noticeable that the characteristic parameters in Eq. (31) are independent of the mode number $l$ because $l$ is an index coupled with the surface deformation $\eta$ to determine the eigenvalues. This is different from the case with a radial forced acceleration, in which the different spatial modes are decoupled and the mode number $l$ can be incorporated to the characteristic parameters in the recursion equation to determine the neutral unstable tongues [32].

The dimensionless quantity $\hat{\gamma} = \gamma/\Omega$ in Eq. (30) is either $0$ or $1/2$, corresponding to the harmonic or subharmonic response of excited surface waves, respectively [29]. Hence, two sets of solutions are obtained. Without loss of generality, since the parameter $m$ is not coupled with $\eta$, we first consider the condition of $m = 0$, which also corresponds to the axi-symmetric problem without emergence of longitudinal modes. This axi-symmetric surface waves can also be observed on the droplet lying on a vibrating plate in many experiments when the amplitude is small [17]

For the condition of $\hat{\gamma} = 0$, $n$ in the recursion relation (30) originates from 1 because the dimensionless parameter $(\Omega/\Omega_c)$ characterizing an experimental condition vanishes if $\hat{\gamma}$ and $n$ are both null. Consider the reality condition for $\eta$, i.e., $\eta_{-1} = \eta_1^*$, where the superscript symbol "*" represents the conjugate of corresponding quantities, the following condition

$$\eta_0(l, m) = \frac{A_0/A_c}{(l-1)(l+2)} \left[ \frac{l-m}{2l-1} \eta_1^r(l-1, m) + \frac{l+m+1}{2l+1} \eta_1^r(l+1, m) \right] \tag{32}$$

shall be satisfied, where $\eta_1^r$ is the real part of $\eta_1$. Eq. (32) indicates that $\eta_0$ is a pure real number for $\hat{\gamma} = 0$. Furthermore, $l$ shall originate from 2 and $\eta_n(l=1) = 0$ since the case $l = 1$ corresponding to the droplet translation is not realized in the present study. Instability boundaries for this problem then can be depicted in the $A_0/A_c - \Omega/\Omega_c$ plane, which are obtained by solving the eigenvalue of the coefficient matrix in Eq. (30). To avoid the singularity of the element in the coefficient matrix, we transformed Eq. (30) into

$$\frac{(l-1)(l+2)l}{(\hat{\gamma}+n)^2}\eta_n^l - \frac{l\,A_0/A_c}{2(\hat{\gamma}+n)^2}\left[\frac{l}{2l-1}(\eta_{n-1}^{l-1}+\eta_{n+1}^{l-1}) + \frac{l+1}{2l+1}(\eta_{n-1}^{l+1}+\eta_{n+1}^{l+1})\right] \quad (33)$$
$$= (\Omega/\Omega_c)^2 \eta_n^l$$

As a result, for each given value of $A_0/A_c$, we can solve a set of $(\Omega/\Omega_c)^2$ that can be recognized as the eigenvalues of the coefficient matrix $\mathbf{M} =$

$$\mathbf{M} \quad (34)$$

Generally, $\mathbf{M}$ is a sparse matrix of infinite dimensions. To solve the eigenvalues of $\mathbf{M}$, we truncate its dimension with a sufficiently large value of $N_d$, and the eigenvalues of lower orders we are interested in are converged numerically **[40]**. In the present study, we truncate $n$ at $N = 5$ and $l$ at $L = 10$, which results in $N_d = N \times (L-1) = 45$ Typically, we are interested in the first subharmonic ($\hat{\gamma} = 1/2$ and $n = 0$) and harmonic ($\hat{\gamma} = 0$ and $n = 1$) responses since most of pervious experimental observations of Faraday instability are limited in these two scenarios **[11, 17, 20, 28]**. The truncation of $n$ at 5 is large enough to obtain the accurate eigenvalues for the present concerns. In addition, after comparing results of $L = 10$ with those of $L = 15$, we can find that the truncation of $l$ at 10 can guarantee the accurate eigenvalues for the mode number $l$ below 8. To get accurate results of higher mode numbers requires larger truncation number of $L$.

The instability boundaries depicted in the $A_0/A_c - \Omega/\Omega_c$ plane for the harmonic response solutions are shown in Fig. 2. In this figure, the instability boundaries for the radial Faraday instability problem **[31]** are also demonstrated for comparison. For each mode of $l$, the unstable tongues originate from $\sqrt{(l-1)(l+2)l}/n$ on the abscissa for both radial and mono-directional acceleration cases. These points indicate the resonance conditions as

$$\left(\frac{\Omega}{\Omega_c}\right)^2 = \frac{(l-1)(l+2)l}{n^2} \text{ or}$$

$$\Omega = \frac{\sqrt{\alpha(l-1)(l+2)l/\rho_1 r_0^3}}{n}, \quad n = 1,2,\ldots \tag{35}$$

for each mode of $l$ in the inviscid limit. This indicates that an infinitesimal forcing acceleration can arise unstable growth of the surface deformation with a specified spherical mode if the forcing frequency satisfies the corresponding condition as defined in Eq. (35).

As shown in Fig. 2, the lateral areas of unstable tongues for both mono-directional and radial acceleration cases are extended with the increase of acceleration amplitude, which means the range of stimulating frequency to arouse the interfacial instability becomes wider with larger acceleration. It is interesting to find that the areas of the unstable tongues for the mono-directional acceleration case shrink greatly compared with those for the radial acceleration case. The unstable frequency range is much narrower for the mono-directional acceleration case at the same forcing amplitude, especially for the cases of small spherical mode numbers.

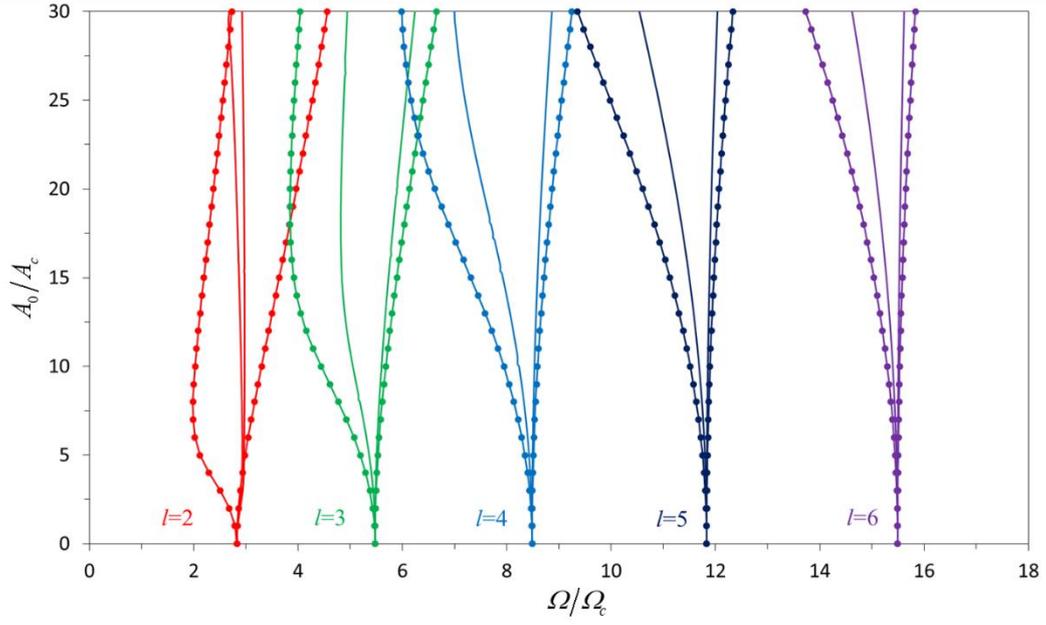

Fig. 2 Harmonic unstable tongues corresponding to the instability with spherical mode numbers $l = 2, 3, 4, 5$ and $6$ for the vertical mono-directional (solid lines) and radial (lines with dots) acceleration cases in the $A_0/A_c - \Omega/\Omega_c$ plane.

For the condition of $\hat{\gamma} = 1/2$, $n$ in the recursion relation (30) originates from 0 and the reality condition for $\eta$, gives $\eta_{-1} = \eta_0^*$, which results in a coefficient matrix $\mathbf{M} =$

$$\begin{pmatrix} 32 & 0 & -1.7143\frac{A_0}{A_c} & 0 & 0 & & 0 & 0 & -1.7143\frac{A_0}{A_c} & 0 & 0 \\ 0 & 32 & 0 & 1.7143\frac{A_0}{A_c} & 0 & & 0 & 0 & 0 & -1.7143\frac{A_0}{A_c} & 0 \\ -3.6\frac{A_0}{A_c} & 0 & 120 & 0 & -2.667\frac{A_0}{A_c} & \cdots 2\times 2(L-1) & -3.6\frac{A_0}{A_c} & 0 & 0 & 0 & -2.667\frac{A_0}{A_c} \cdots (N+1)\times 2(L-1) \\ 0 & 3.6\frac{A_0}{A_c} & 0 & 120 & 0 & & 0 & -3.6\frac{A_0}{A_c} & 0 & 0 & 0 \\ 0 & 0 & -4.5714\frac{A_0}{A_c} & 0 & 288 & & 0 & 0 & -4.5714\frac{A_0}{A_c} & 0 & 0 \\ & \vdots 2\times 2(L-1) & & & & \ddots & & \vdots & & & \cdots \\ 0 & 0 & -0.1905\frac{A_0}{A_c} & 0 & 0 & & & & & & \\ 0 & 0 & 0 & -0.1905\frac{A_0}{A_c} & 0 & & 3.5556 & 0 & 0 & 0 & 0 \\ & & & & & & 0 & 3.5556 & 0 & 0 & 0 \\ -0.4\frac{A_0}{A_c} & 0 & 0 & 0 & -0.2963\frac{A_0}{A_c} & \cdots & 0 & 0 & 13.333 & 0 & 0 & \cdots \\ 0 & -0.4\frac{A_0}{A_c} & 0 & 0 & 0 & & 0 & 0 & 0 & 13.333 & 0 \\ 0 & 0 & -0.5079\frac{A_0}{A_c} & 0 & 0 & & 0 & 0 & 0 & 0 & 32 \\ & \vdots (N+1)\times 2(L-1) & & \vdots & & \vdots & & & & \ddots & \end{pmatrix} \quad (36)$$

The approach to solve the eigenvalues of Eq. (36) to get the instability tongue for subharmonic response is the same as Eq. (34), which are shown in Fig. 3. For the case of $A_0/A_c = 0$, we can easily get the resonance conditions as

$$\left(\frac{\Omega}{\Omega_c}\right)^2 = \frac{4(l-1)(l+2)l}{(2n+1)^2} \quad \text{or} \tag{37}$$

$$\Omega = \frac{2\sqrt{\alpha(l-1)(l+2)l/\rho_1 r_0^3}}{2n+1}, \quad n = 0,1,2,\dots$$

for the subharmonic oscillation, which stand for the intersection points on the abscissa in Fig. 3 for different spherical mode numbers $l$.

The subharmonic unstable tongues for the radial Faraday instability problem are also depicted in Fig. 3 as the lines with dot marks for comparison. It is found that for the mono-directional acceleration case, the unstable tongues converge to a set of straight lines parallel with the ordinate. That is to say, the points in the unstable regions for the radial acceleration case but outside the straight lines become stabilized for the mono-directional acceleration case, even though the amplitude and frequency of the vibration are kept the same. This theoretical result is to be validated by the experiments as discussed in §5.

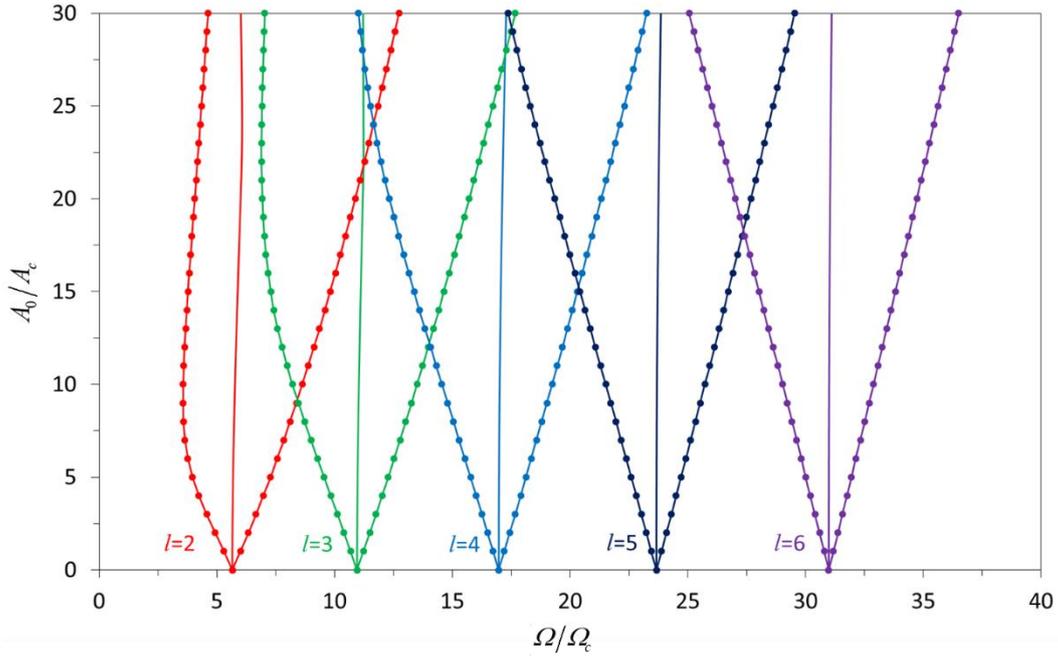

Fig. 3 Subharmonic unstable tongues corresponding to the instability with spherical wavenumbers $l = 2, 3, 4, 5$ and $6$ for the vertical mono-directional (solid lines) and radial (lines with dots) acceleration cases in the $A_0/A_c - \Omega/\Omega_c$ plane.

The effect of the longitudinal mode number $m$ is not intrinsic because $m$ is not coupled with $\eta$. Fig. 4 shows the harmonic unstable tongues ($\hat{\gamma} = 0$) for different $m$-values with fixed mode numbers $l = 4$ and $l = 5$. As $m$-value increases, the right boundary of each tongue is merely changed while the left one moves rightwards, which shrinks the region of unstable tongue gradually. Particularly, the unstable tongue for the condition $m = l$ (sectorial mode) almost shrinks to a line, indicating that the pure longitudinal unstable waves are hardly realized on the droplet surface.

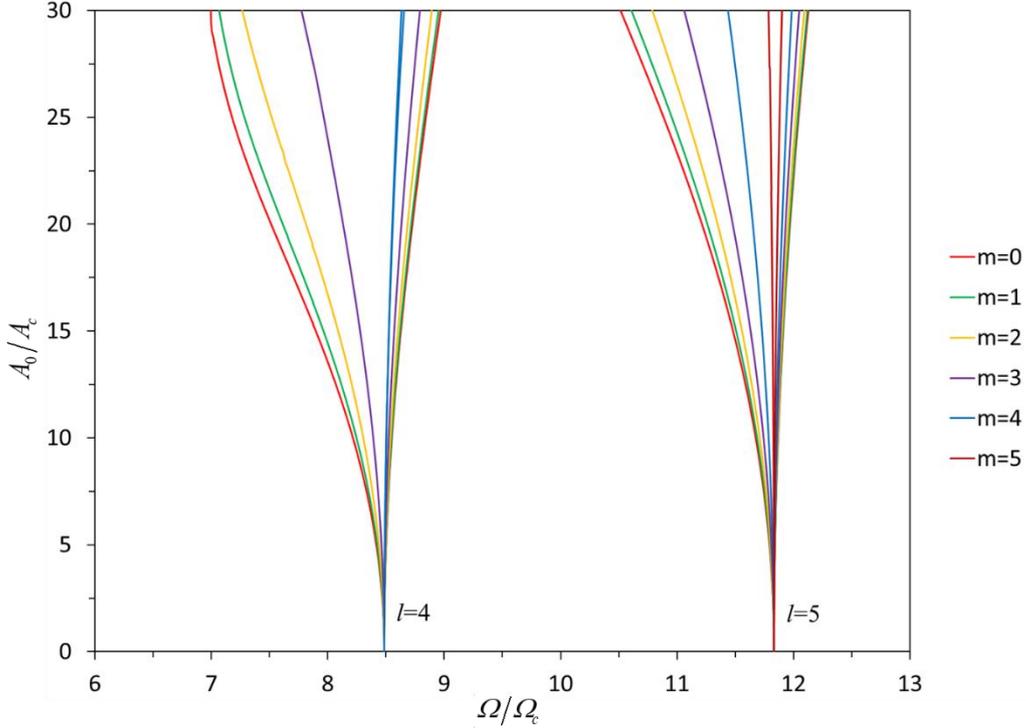

Fig. 4 Harmonic unstable tongues for different $m$-values with $l = 4$ and $l = 5$.

## 5 Experimental observations on the unstable surface waves of liquid droplets

To verify the theoretical prediction of the harmonic response on the spherical surface under vertical mono-directional acceleration in § 4, the temporal evolutions of the excited unstable surface waves on a quasi-semi-spherical droplet lying on a vertically vibrating plate were experimentally recorded, with the focus on the oscillating period of the surface waves.

Fig 5 shows the experimental system in which the surface deformation of a droplet lying on a vertically vibrating plate can be traced optically with independent control of frequency and amplitude. The experimental system mainly includes: the piezoelectric ceramic sheet, support device, control unit and optical test unit. The plate was connected with the upper end of the JZK2 Vibration Exciter and would vibrate at the same frequency as sinusoidal voltage input in the vertical direction. UTG 9005C Function Signal Generator was used to provide sinusoidal wave signal with excellent stability. In the experiment, the frequency was set to be ranging from 100 Hz to 500 Hz with an interval of 50 Hz. SA-PA010 power amplifier was coupled with signal generator to provide excitation signal. The frequency and voltage of signal were shown by Ds1054 Oscilloscope. The amplitude of the vibrating plate is adjusted by varying different excitation voltage. The syringe was used to produce deionized water droplets with diameters of $7.5 \pm 0.05$ mm.

MacroVis EoSens high-speed digital camera produced by HSVISION coupled with Tamron 180 mm focal length lens was used to take a single shot. The pixel resolution was $512 \times 512$, the frame rate was 11000 fps and exposure time was 88 μs. The dysprosium lamp with the highest power of 2500 W was placed on the opposite side of the camera to provide the light source. The motion of the plate and the deformation of surface wave on the droplet were recorded by shadowgraph technique.

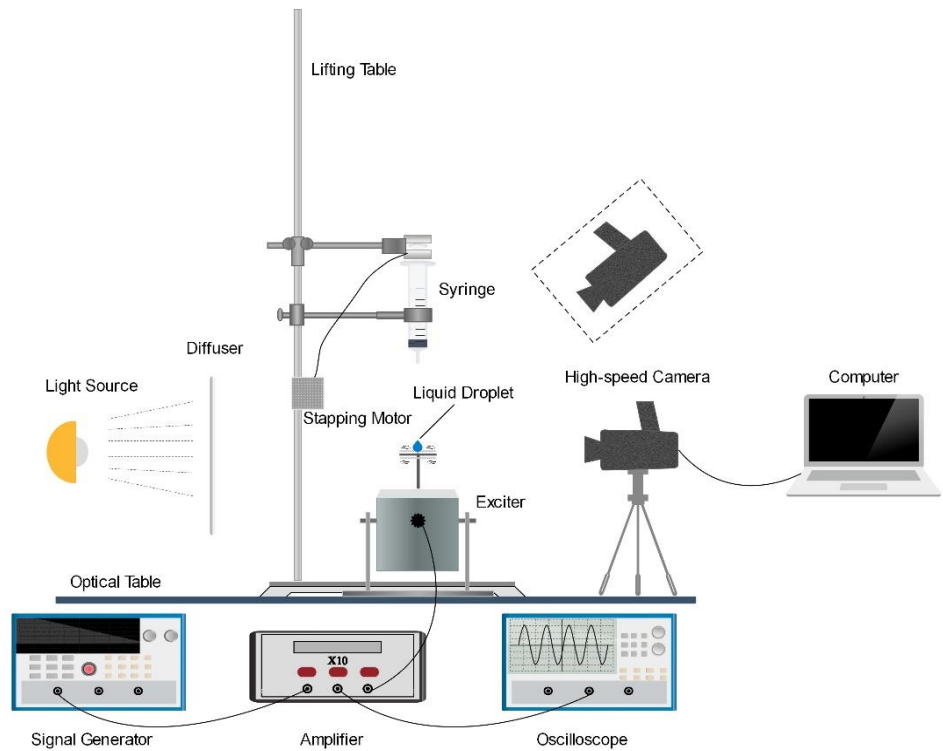

Fig. 5 Schematic of the experimental setup for the surface response of a liquid droplet subject to vertical vibration.

Fig. 6 shows the typical deformation of droplet surface for different voltages with the plate vibration frequency being fixed at 300 Hz. After an initial transient stage, the droplet surface enters a steady deformation stage, during which different dynamic behaviors can be observed: (*i*) when the amplitude of the vibrating plate is small at excitation voltage of 5.7 V as shown in Fig. 6 (b), only latitudinal waves emerges on the droplet surface, so we can recognize that the longitudinal mode number $m$ is equal to zero for this case; (*ii*) in the case of 11.1 V in Fig. 6 (c), the droplet surface shows distinct latitudinal modes and some longitudinal waves appear at the bottom part of the droplet; (*iii*) in the case of 20 V in Fig. 6 (d), both the latitudinal and longitudinal waves begin to interfere with each other and (*iv*) complicated structures composed of more large wave modes are observed on the droplet surface under the mutual interference of these two kind of waves in the case of 25 V can be observed in Fig. 6 (e).

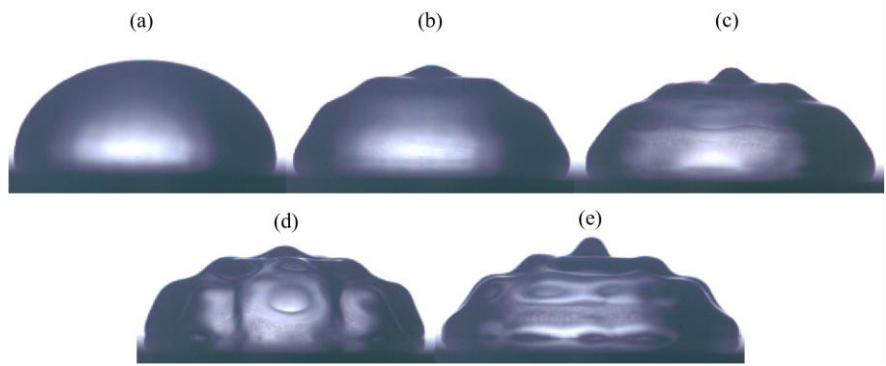

Fig. 6 Snapshots of surface deformation of liquid droplet (diameter ≈ 7.5 mm) on a vibrating plate forced by a sinusoidal excitation for different conditions with the voltages of (a) 0 V, (b) 5.7 V, (c) 11.1 V, (d) 20 V and (e) 25 V. The snapshots of (b) to (e) correspond to the top dead point of vibration in steady deformation state. The frequencies for all vibration cases (b) to (e) are fixed at 300 Hz.

Fig.7 shows the temporal evolution of the droplet surface in steady deformation stage at frequency of 300 Hz and voltage of 5.7 V. To ensure that the droplet is in steady state at this voltage, it is necessary to slowly increase the excitation voltage to 5.7 V and remain it for sufficient long time. The snapshots of the surface deformation shape at five representative moments, *i.e.*, t = $0T, 0.25T, 0.5T, 0.75T, 1T$, where $T = 1/f$ is the vibration period, are demonstrated in Fig. 7 and the focus is on the period of the excited surface wave.

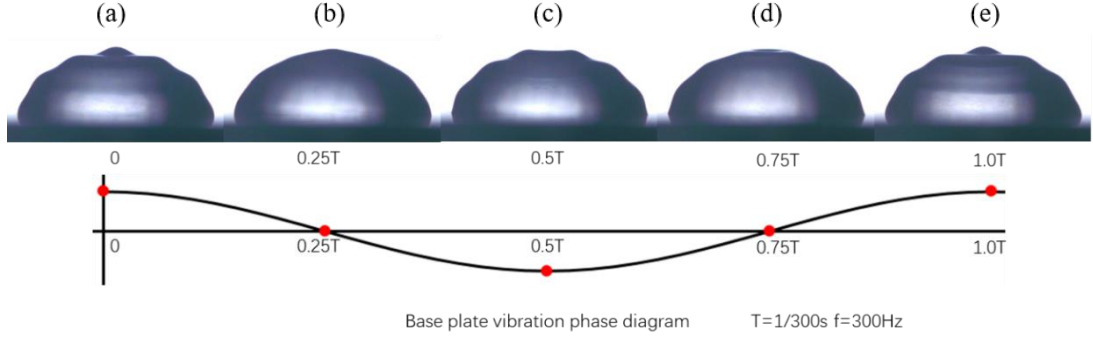

Fig. 7 Temporal evolution of the droplet surface deformation during one vibration period for the case of $f = 300$ Hz and $U = 5.7$ V. The snapshots correspond to (a) $t/T = 0$, (b) $t/T = 0.25$, (c) $t/T = 0.5$, (d) $t/T = 0.75$, (e) $t/T = 1$, where $t$ is the time and $T = 1/f$ is the vibration period.

As shown in Fig.7, at this frequency and voltage ($f = 300$ Hz and $U = 5.7$ V), the droplet surface waves only exhibit the latitudinal waves ($m = 0$). The temporal evolutions of the surface deformation clearly show that the frequency of the surface wave is equal to that of the vibrating plate. This observation confirms that the excited unstable surface wave on the droplet is indeed the harmonic response. For other cases with different frequencies and voltages, we also found that the frequencies of droplet surface waves are all equal to those of vibrating plate when only latitudinal waves emerge on the droplet surface. This phenomenon is different from the theoretical predictions on the Faraday instability in planar liquid layers and the spherical droplets with radial acceleration, in which the subharmonic response of the unstable surface wave is more prominent **[28, 32]**.

In fact, some previous researches have also reported the harmonic response of the axisymmetric standing waves excited on the free surface of the vertically vibrating droplet present at small values of the excitation amplitude **[17, 23]**, but there lack sufficient physical interpretations. The theoretical results derived in §4 can well explain why the latitudinal waves ($m = 0$) emerging on the droplet surface are always harmonic under vertical vibration.

In §4, we compared the similarities and differences between the mono-directional and radial vibrating droplets by linear analysis. We found that although their unstable tongues have the same intersection points with the abscissa axis, the areas of the unstable tongues for the mono-directional acceleration case shrink greatly compared with those for the radial acceleration case. This indicates that the mono-directional vibrating droplet is less easily destabilized than radial vibrating droplet.

The most remarkable difference between the mono-directional and radial vibrating droplets lies in the subharmonic case. The area of subharmonic tongues for the radial vibrating droplet is rather large, while the left and right boundary lines of the subharmonic unstable tongue for the mono-directional vibrating droplet almost coincide with each other, which means that the subharmonic unstable modes can hardly be excited in the mono-directional vibration case. According to the theoretical analyses of the radial vibrating droplet, when the amplitude is slowly increased with a fixed frequency, the first emerging droplet surface wave mode should be subharmonic **[32]**. However, no matter how slowly we increase the amplitude in the experiments, the first appearing unstable latitudinal waves on the droplet surface are always harmonic at all frequencies until the emergence of longitudinal waves. This is well predicted by the theoretical analysis on the mono-directional vibration case in §4.

## 6. Conclusions

The interfacial instability of a spherical droplet subject to a vertical mono-directional vibration was theoretically investigated, which yielded different outcomes and dynamic behaviors of the destabilized surface from those derived for the spherical instability problem with radial vibration.

The linearized governing equations and corresponding boundary conditions for the spherical droplet under vertical vibration were derived and solved by Floquet analysis. Similar to the classic Faraday instability for the liquid layer and radial-vibrating droplet, a recursion equation that implicitly expresses the dispersion relation between the growth rate and spherical mode numbers was obtained. However, the mono-directional vibration results in a polar-angle dependent acceleration in radial direction acting on the droplet in the spherical coordinate. As a result, the Fourier components of the surface deformation displacement from different spherical harmonics are closely coupled in the recursion equation. This is intrinsically different from the case of droplets under radial vibration, in which the spherical mode numbers can be treated separately.

For the inviscid condition, solving the positive eigenvalues of the coefficient matrix for $\hat{\gamma} = 1$ and $1/2$, we obtained neutral stable boundaries for the harmonic and subharmonic responses of excited surface waves in the $A_0/A_c - \Omega/\Omega_c$ parameter plane. For the harmonic response, the unstable tongues of the mono-directional acceleration case shrink greatly compared with those under radial acceleration at the same spherical mode number. For the subharmonic response, the unstable tongues of the mono-directional acceleration case converge to a set of straight lines parallel with the ordinate, which indicates that the unstable spherical modes in the radial acceleration case are hardly destabilized in the mono-directional acceleration case. This theoretical prediction is then verified experimentally by lying a water droplet on a vertically vibrating plate. As the value of the longitudinal mode number $m$ increases, the right boundary of each tongue in the parameter plane is merely changed while the left one moves rightwards, which shrinks the region of unstable tongue gradually.

## Acknowledgement

The work was supported by the National Natural Science Foundation of China (Grant No. 51976011), Beijing Natural Science Foundation (Grant No. 3212022) and State Key Laboratory of Engines, Tianjin University (K2020-02).

## Declaration of Interests

The authors report no conflict of interest.